\documentclass[final,authoryear,3p,times,twocolumn]{elsarticle}
\usepackage{graphicx}
\usepackage{epsfig}
\usepackage{amssymb}

\journal{New Astronomy}
\begin{document}
\begin{frontmatter}
\title{The 2015 super-active state of recurrent nova T CrB and the long term evolution after the 1946 outburst}

\author[un,um]{Ulisse Munari} 
\author[an]{Sergio Dallaporta} 
\author[an]{Giulio Cherini}

\address[un]{corresponding author: Tel.: +39-0424-600033, Fax.:+39-0424-600023, e-mail: ulisse.munari@oapd.inaf.it}
\address[um]{INAF Astronomical Observatory of Padova, via dell'Osservatorio 8, 36012 Asiago (VI), Italy}
\address[an]{ANS Collaboration, c/o Astronomical Observatory, 36012 Asiago (VI), Italy}

\begin{abstract}
The recurrent nova T CrB has entered in 2015 a phase of unprecedented high
activity.  To trace something equivalent, it is necessary to go back to
1938, before the last nova eruption in 1946.  The 2015 super-active state is
characterized by: a large increase in the mean brightness ($\Delta$$B$=0.72
mag over the uderlying secular trend), vanishing of the orbital modulation
from the $B$-band lightcurve, and appearance of strong and high ionization
emission lines, on top of a nebular continuum that overwhelms at optical
wavelengths the absoption spectrum of the M giant.  Among the emission
lines, HeII 4686 attains a flux in excess of H$\gamma$, the full set of OIII
and NIII lines involved in the Bowen fluorescence mechanism are strong and
varying in intensity in phase with HeII 4686, and OIV and [NeV] are present. 
A large increase in the radiation output from the hot source is reponsible for a
large expansion in the ionized fraction of the M giant wind.  The wind is
completely ionized in the direction to the observer.  A high electron
density is supported by the weakness of forbidden lines and by the large
amplitude and short time scale of the reprocessing by the nebular material
of the highly variable photo-ionization input from the hot source.  During the
super-active state the nebula is varying to and from ionization-bounded and
density-bounded conditions, and the augmented irradiation of the cool giant
has changed the spectral type of its side facing the hot source from M3III to M2III,
i.e.  an increase of $\sim$80~K in effective temperature.
\end{abstract} 

\end{frontmatter}

\section{Introduction}
\label{}

T CrB is one of the few known Galactic recurrent novae (Warner 1995,
Schaefer 2010), with outbursts recorded in 1866 and 1946.  They have been
spectacular events, peaking around 2.0 mag (Pettit 1946), displaying nearly
identical lightcurves characterized by an extremely fast rise to maximum and
a rapid decline, taking $t_2$=3.8 days to decline by 2 mag (Payne-Gaposchkin
1957).  The spectroscopic evolution (Sanford 1946, 1949) was - in modern
terms - that of an He/N nova (Williams 1992), reaching very high
ionization conditions as indicated by the presence of strong [FeX] and
[FeXIV] coronal lines (Sanford 1947).  Peculiar to T CrB is the presence on
both outbursts, of a secondary, fainter and broader maximum $\sim$110 days
past the primary maximum, which physical nature is still debated (e.g.
Webbink 1976, Cannizzo \& Kenyon 1992, Selvelli, Cassatella, \& Gilmozzi
1992, Ruffert, Cannizzo, \& Kenyon 1993).

The donor star in T CrB is an M3III, filling its Roche lobe (Bailey 1975,
Yudin \& Munari 1993), on a 227.55 day orbit (Kenyon \& Garcia 1986, Fekel
et al.  2000) around a WD companion (Selvelli, Cassatella, \& Gilmozzi 1992,
Belczynski \& Mikolajewska 1998).  The presence of a cool giant makes T CrB
also a member of the class of symbiotic binaries (Allen 1984, Kenyon 1986),
similarly to the other symbiotic recurrent novae RS Oph, V745 Sco and V3890
Sgr (Munari 1997).

The brightness in quiescence ($V$$\sim$10 mag) and the favourable position
on the sky (absence of seasonal gaps in the observability from the northern
emisphere), have fostered continued interest and study of T CrB. In
quiescence, typical symbiotic binaries display a rich and high ionization
emission line spectrum (comprising [NeV], [CaV], [FeVII], HeII, and Raman
scattered OVI), superimposed to the absorption spectrum of the cool giant,
with the nebular continuum veiling in the yellow, and overwhelming in the
blue, its molecular absorption bands (Allen 1984, Munari and Zwitter 2002,
Skopal 2005).  As a symbiotic binary, the optical quiescence spectrum of T
CrB is atypical in showing very little else than the M3III absorption
spectrum.  Most of the time, only a weak emission in H$\alpha$ is noticeable
on low resolution spectra (Kenyon 1986).  On rare occasions, a surge in
activity causes the optical spectrum of T CrB to show something more typical
of symbiotic binaries, e.g. Balmer lines and continuum in emission, and
sometimes even the appearance of a weak HeII 4686 in emission (Iijima 1990,
Anupama \& Prabhu 1991).

In this paper we report on the super-active conditions displayed by T CrB
during 2015 (in the following SACT-2015 for short), conditions never seen
before. SACT-2015 appears to be much stronger, both photometrically and
spectroscopically, than previous periods of enhanced activity recorded after
the 1946 nova outburst.

   \begin{table*}
      \caption{Our $B$$V$$R_{\rm C}$$I_{\rm C}$ photometric observations of T CrB.
      The full table is available electronically via CDS, a small portion is
      shown here for guidance on its form and content.}
       \centering
       \includegraphics[width=120mm]{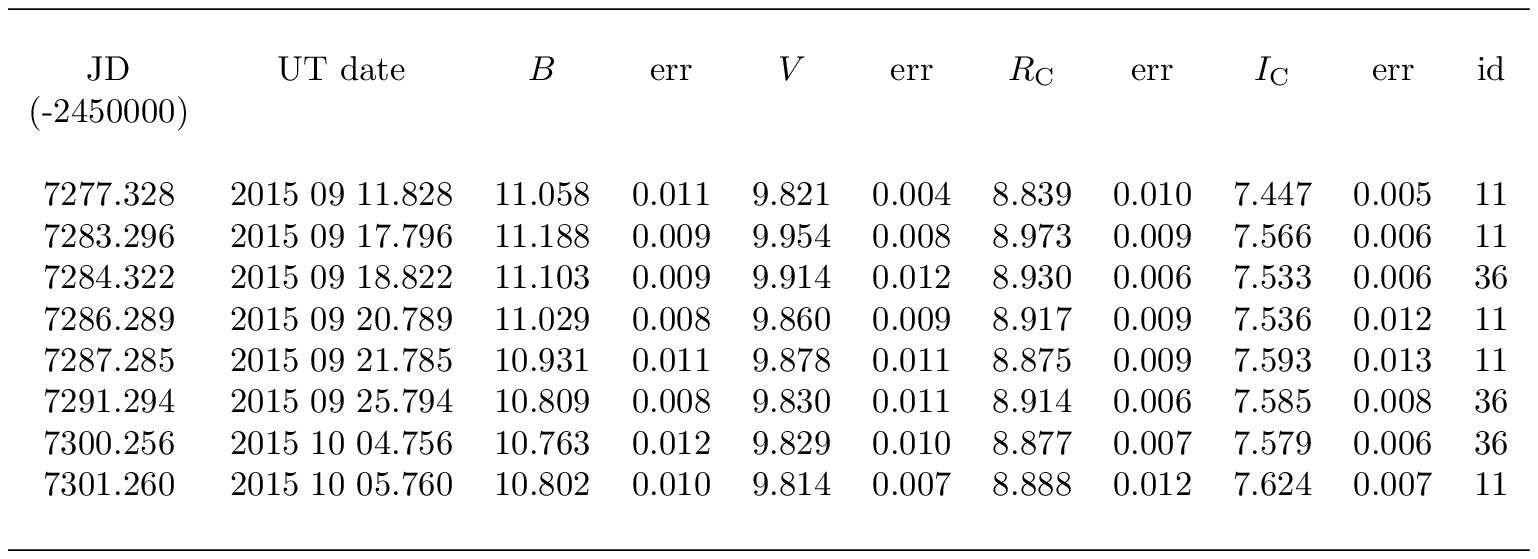}
       \label{tab1}
   \end{table*}

\section{Observations}

$B$$V$$R_{\rm C}$$I_{\rm C}$ optical photometry of T CrB is regularly obtained since
2006 with ANS Collaboration telescopes N.  11 and 36, located in Italy in
Trieste and Cembra, respectively.  The star has been observed on 205 nights,
from May 11, 2006 to Dec 20, 2015.  The operation of ANS Collaboration
telescopes is described in detail by Munari et al.  (2012) and Munari \&
Moretti (2012).  The same local photometric sequence, calibrated by Henden
\& Munari (2006) against Landolt equatorial standards, was used at both
telescopes on all observing epochs, ensuing a high consistency of the data.  
The $B$$V$$R_{\rm C}$$I_{\rm C}$ photometry of T CrB
is given in Table~1, where the quoted uncertainties are the total error
budget, which quadratically combines the measurement error on the variable
with the error associated to the transformation from the local to the
standard photometric system (as defined by the photometric comparison
sequence).  All measurements were carried out with aperture photometry, the
long focal length of the telescopes and the absence of nearby contaminating
stars not requiring to revert to PSF-fitting.

Low resolution spectra of T CrB were obtained with the 1.22m
telescope + B\&C spectrograph operated in Asiago by the Department of
Physics and Astronomy of the University of Padova.  The CCD camera is a
ANDOR iDus DU440A with a back-illuminated E2V 42-10 sensor, 2048$\times$512
array of 13.5 $\mu$m pixels.  It is highly efficient in the blue down to the
atmospheric cut-off around 3250~\AA. The spectral dispersion is 2.31 Ang/pix
and the spectral resolution is constant at $\sim$2.2 pix, with the spectra
extending from $\sim$3300 to $\sim$8050~\AA. The slit width has been kept
fixed at 2 arcsec, and the slit always alligned with the parallactic angle
for optimal absolute flux calibration.

High resolution spectra were obtained with the Echelle spectrograph mounted
on the 1.82m Asiago telescope.  It is equipped with an EEV~CCD47-10 CCD,
1024$\times$1024 array, 13 $\mu$m pixel, covering the interval
$\lambda\lambda$~3600$-$7300~\AA\ in 32 orders, at a resolving power of
20\,000 and without inter-order wavelength gaps.

  \begin{figure*}[!Ht]
     \centering
     \includegraphics[width=16cm]{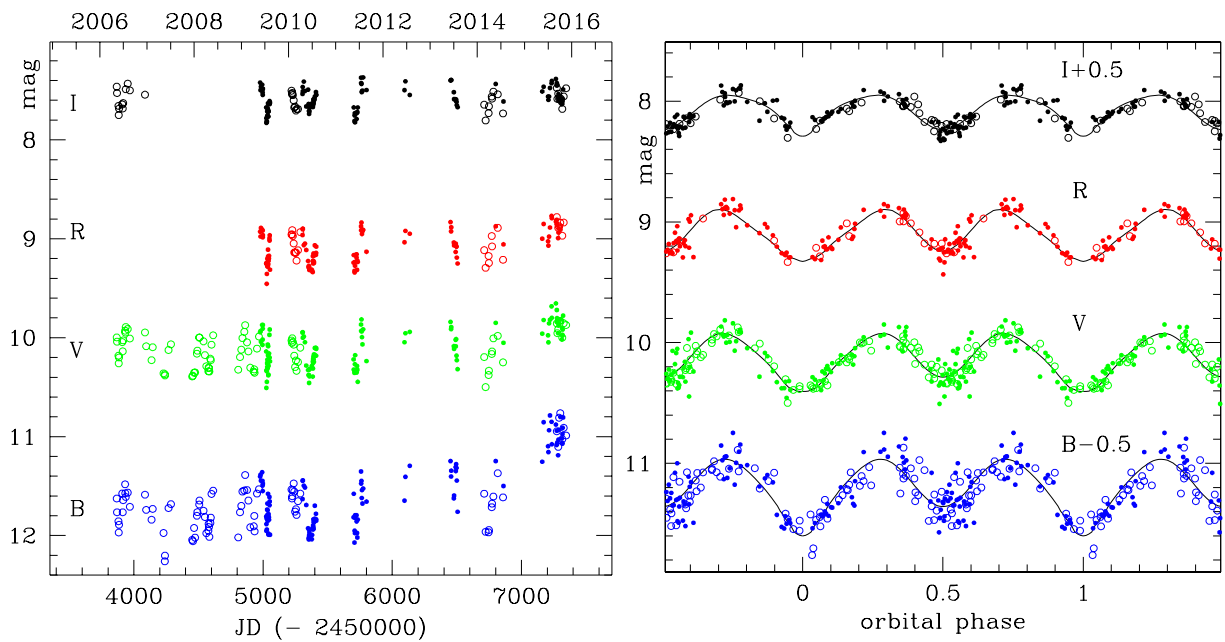}
     \caption{{\em Left}: the 2006-2015 $B$$V$$R_{\rm C}$$I_{\rm C}$
     lightcurves of T CrB based on our data in Table~1 (solid and open
     circles mark observations obtained with ANS Collaboration telescopes 11
     and 36, respectively). The surge in brightness during 2015 is
     prominent. {\em Right}: the quiescence part of the data at left is here
     phase plotted against the P=227.55 days orbital period.  The curves are
     low order Legendre polynomials, symmetric with respect to the WD
     transit at lower conjunction (phase 0.5), to provide a simple fit to
     guide the eye to the ellipsoidal modulation.}
     \label{fig1}
  \end{figure*}

\section{Photometric evolution during 2006-2015}

The 2006-2015 $B$$V$$R_{\rm C}$$I_{\rm C}$ lightcurves of T CrB from our CCD
observations of Table~1 are plotted on the left panel of Figure~1.  The much
brighter state of T CrB in 2015 is evident, with increasing revelance toward
shorter wavelengths.  On the right panel of Figure~1, the 2006-2014 data
(preceeding SACT-2015) are phase plotted against the orbital ephemeris
\begin{equation}
{\rm Min} I = 2431933.83 + 227.55 \times E  
\end{equation}
which gives the epochs of primary minima (passages of the M3III companion at
inferior conjunction) for the orbital period derived by Kenyon \& Garcia
(1986) and Fekel et al.  (2000).  The resulting phased lightcurve is
dominated by the well known ellipsoidal distortion of the M3III giant, first
reported by Bailey (1975) at optical wavelengths and by Yudin \& Munari
(1993) in the infrared.  The over-plotted curves are simple fits to guide the
eye, in particular to demonstrate that: ($a$) the amplitude of
the ellipsoidal modulation, as given by the fitting curves, is $\Delta
B$=0.63, $\Delta V$=0.49, $\Delta R_{\rm C}$=0.42, and $\Delta I_{\rm
C}$=0.34 mag; ($b$) the secondary minimum (phase 0.5, WD passing at inferior
conjunction) is shallower at shorter wavelengths, a fact due to
the irradiation of the cool giant by the hot WD companion; and ($c$)
the dispersion of points around the fitting curves increases toward shorter
wavelengths and is well in excess of the small observational errors (cf
Table~1).  The reason for that is associated with the erratic
behaviour of accretion phenomena in the system.  In fact, a long record of
observations document a large amplitude {\em flickering} affecting time series
observations of T CrB (eg.  Zamanov \& Bruch 1998, Zamanov et al.  2004,
Gromadzki et al.  2006, Dobrotka et al.  2010).

  \begin{figure}[!Ht]
     \centering
     \includegraphics[width=7.8cm]{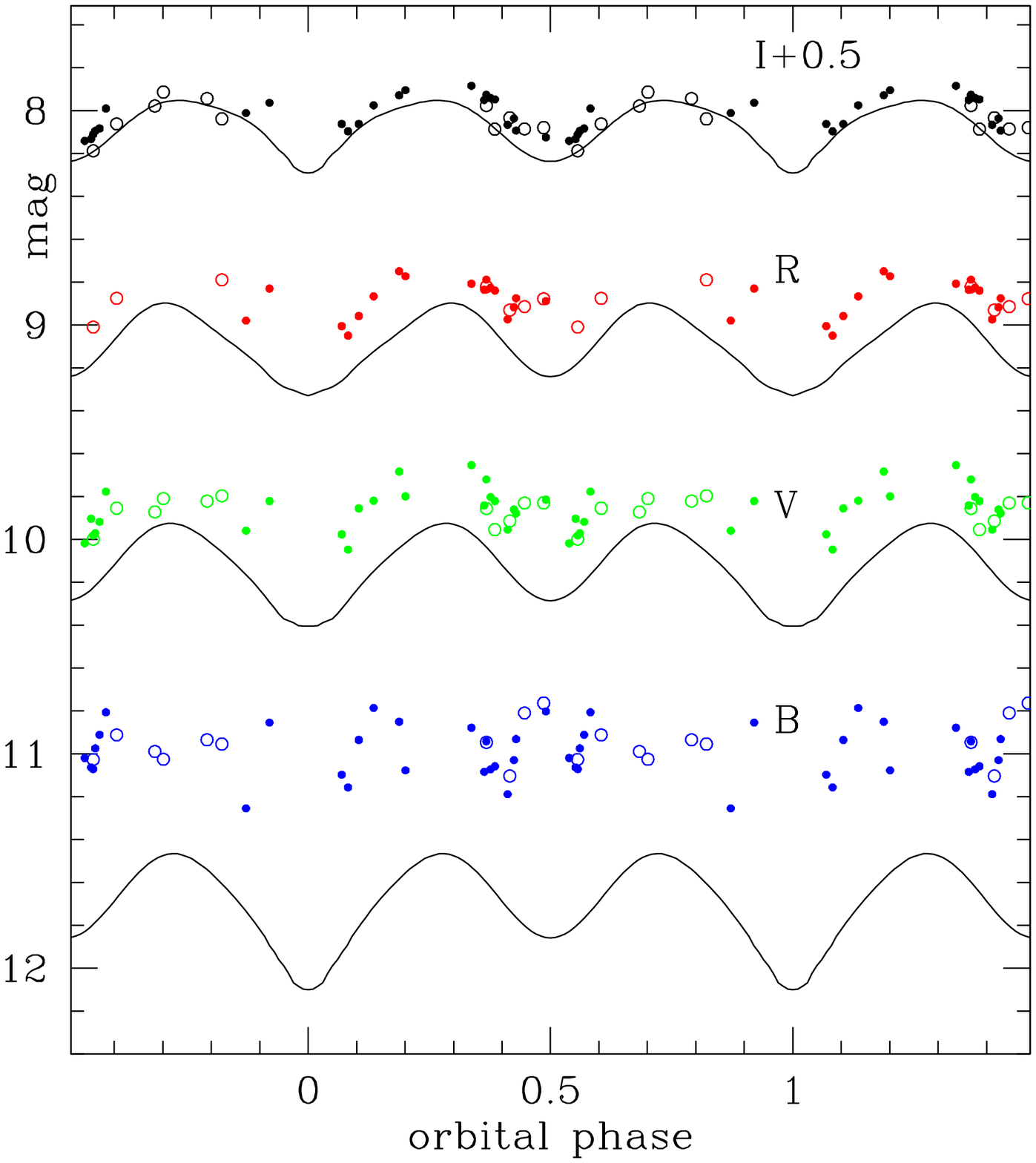}
     \caption{The photometric observations of T CrB during the 2015 super-active 
     period are phase-plotted against the orbital ephemeris (Eq. 1), with
     mean curves for quiescence imported from Figure~2.}
     \label{fig2}
  \end{figure}

The photometric data on T CrB secured during SACT-2015 are not inserted in
the right panel of Figure~1, which only deals with the preceding quiescence. 
The SACT-2015 data are instead plotted in Figure~2, against the same
ephemeris as used for Figure~1, from which the polynomial fits are also
copied.  From Figure~2 we infer that: (i) the shorter the wavelength, the
larger the increase in brightness during SACT-2015: compared with preceding
quiescence (as given by the fitting curves in Figures~1 and 2), the increase
is $\Delta$$B$=0.72, $\Delta$$V$=0.28, $\Delta$$R_{\rm C}$=0.21, and
$\Delta$$I_{\rm C}$=0.09 mag; (ii) during SACT-2015, the orbital modulation
disappears from the $B$ lightcurve, and the depth of secondary minimum
(orbital phase 0.5) is reduced in the $V$ and $R_{\rm C}$ lightcurves; and
(iii) contrary to previous high states that did not influence T CrB
brightness at longer wavelengths, the effect of SACT-2015 extends well into
the far red, with all $I_{\rm C}$ measurements laying above the polynomial
fit to quiescence data.  Anticipating the spectroscopic results from
following sections, these photometric signatures are due to: (1) nebular emission from
a much larger
fraction of the M3III wind now ionized by the hot source, which visibility is not
affected by orbital motion, and (2) increased irradiation and
therefore higher re-emission from the side of the M3III facing the hot source.  In
this respect it is interesting to note that the dispersion of the
observations is similar during SACT-2015 and quiescence, with
$\sigma$($B$)=0.115 and 0.120 mag, respectively.  This suggests that the
increased output from the hot source is powered by the same accretion processes and
associated instabilities that dominates during quiescence.  In addition, the
large amplitude and short time scale (of the order of a day or a few days at
most) of these erratic fluctuations, suggests that the electron density in
the ionized gas (dominating the system brightness in $B$) is high enough to
drive a short recombination time scale, so that the short time scale and
large amplitude of the variations in the photo-ionization input are not
washed out by reprocessing from the recombining gas.

\section{Long term 1947-2015 evolution and previous active states}

To place SACT-2015 into a broader perspective, we investigated the long term
brightness evolution of T CrB following the 1946 nova outburst.  The
outburst was over by the summer of 1947 when the star had returned to
quiescence brightness, $V$$\sim$10 mag. This value is similar to what
F.~W.~A.  Argelander measured in 1855 - before the 1866 nova outburst - for
his Bonner Durchmusterung (BD) star atlas, and to the brighntess that
characterized the star in between the two nova eruptions of 1866 and 1946
(Barnard 1907, Campbell \& Shapley 1923).

To reconstruct the 1947-2015 lightcurve of T CrB, we selected to use the
120,000 visual estimates collected by AAVSO (privately communicated by
Stella Kafka, Director).  This choice is based on two primary reasons.

The first in that the visual estimates collected by AAVSO constitute an
uninterrupted record of T CrB brightness during the last 70 years, whereas
other sources of information (measurements in any photometric band) are too
sparse in time and scattered through so many different observers,
instrumental combinations and photometric systems to be of no use to our
goal.

The second argument is based on the fact that during the last 70 years, each
epoch has been chacarterized by a different type of instrument to record
stellar brightness: initially unfiltered blue sensitive photographic
emulsions, then filtered pancromatic photographic emulsions, followed by
photoelectric photometers, and finally by CCD devices.  The very red color
of T CrB (much redder than most of the suitable comparison stars around the
variable) has impacted in different ways and by different amounts the
photometry collected with such a broad assortment of instruments.  In
addition, only rarely the data have been properly transformed to standard
systems, most of the measurements being just differential with respect to a
single field star (usually of unmatching colors).  On the contrary, visual
estimates seem to be far more stable over different epochs and subsequent
generations of observers: the comparison sequence has not changed much and
the many different observers participating in the 70 years of AAVSO
monitoring have used the same measuring device, their unfiltered eyes.

   \begin{table}[!Ht]
      \caption{Integrated fluxes (in units of 10$^{-13}$ erg cm$^{-2}$
      sec$^{-1}$) of emission lines in the spectrum of
      T CrB for 2015-10-16 from Figure~4.}
       \centering
       \includegraphics[height=165mm]{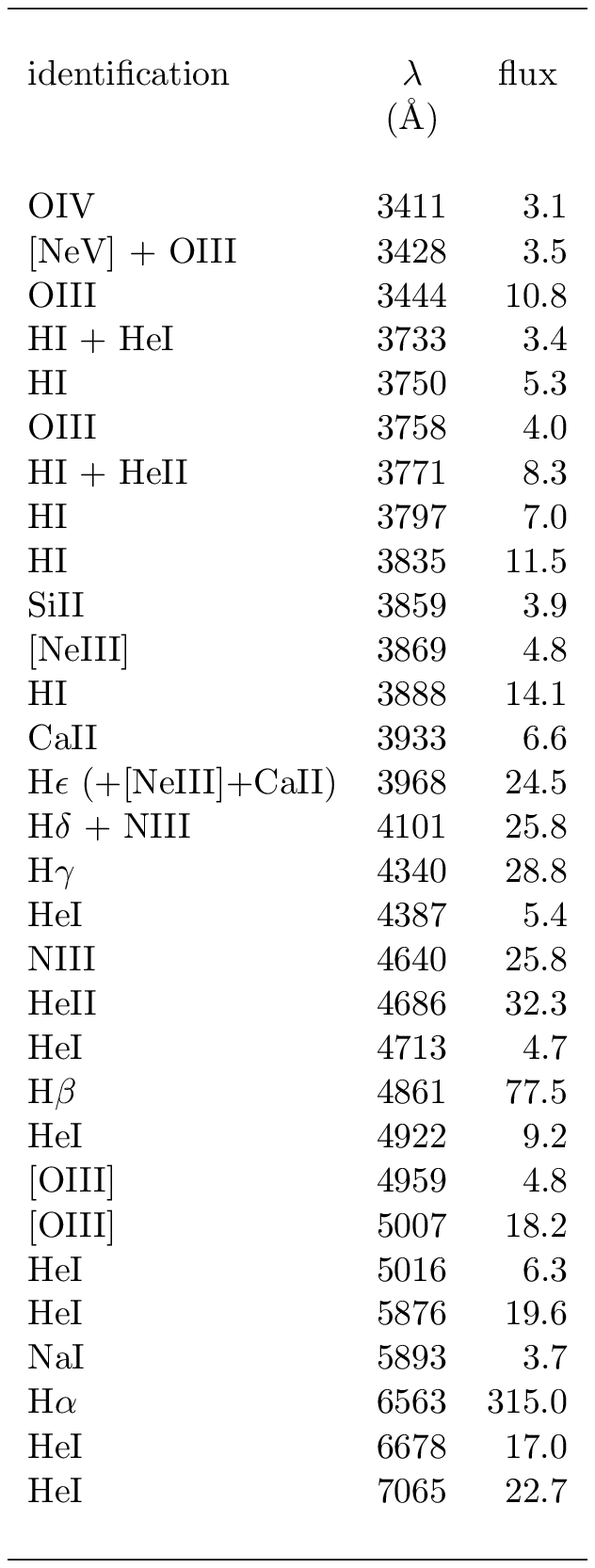}
       \label{tab2}
   \end{table}

\subsection{Mean magnitudes}

In tracking the secular evolution of T CrB following the 1946 outburst, we
are looking for subtle effects (of the order of hundredths of a magnitude),
much less than the scatter intrinsic to visual estimates.  We have to filter
out the noise.

  \begin{figure}[!Ht]
     \centering
     \includegraphics[width=7.5cm]{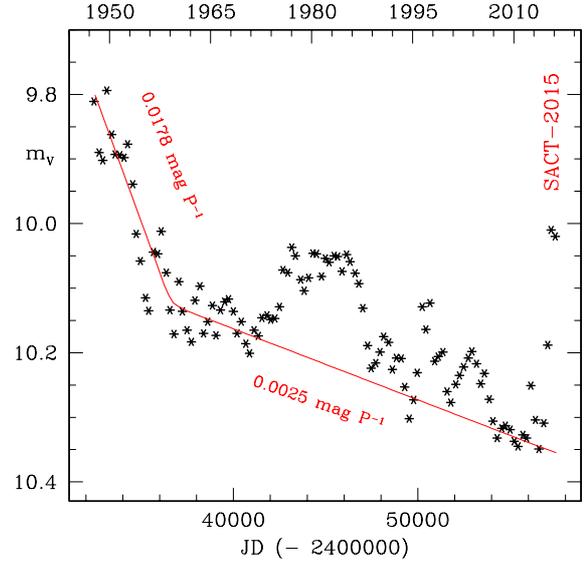}
     \caption{Long term evolution of the brightness of T CrB in quiescence,
     from 120,000 AAVSO visual estimates collected after the 1946 nova
     outburst. Each point represents the mean value of the visual estimates
     covering an orbital cycle (227.55 days).}
     \label{fig3}
  \end{figure}

The simplest way could be to average all data within a given time step. Such
a straight average would however most likely generate spurious signals.  In
fact, the majority of the visual estimates are concentrated during the
summer and autumn months, when T CrB is best located in the evening sky. 
Given the long orbital period of the system (about 62\% of 1 year), this
would mean that in different years the system is on average observed at
different orbital phases.  Given the large amplitude of the orbital
modulation, this would cause a spurious beating signal.

To overcome the problem, we have divided the AAVSO data into contingous
orbital cycles (110 cycles from 1947 to current time), and $\chi^2$ fitted
to them the average phased lightcurve for quiescence (the continous curve
for $V$ band in Figure~1).  In this way, no matter how unevenly distributed
in orbital phase the observations could be, a correct estimate for the
magnitude averaged along a whole orbital cycle is obtained.  The values so
derived are plotted in Figure~3.

\subsection{The secular trend and past active phases}

  \begin{figure*}[!Ht]
     \centering
     \includegraphics[width=16cm]{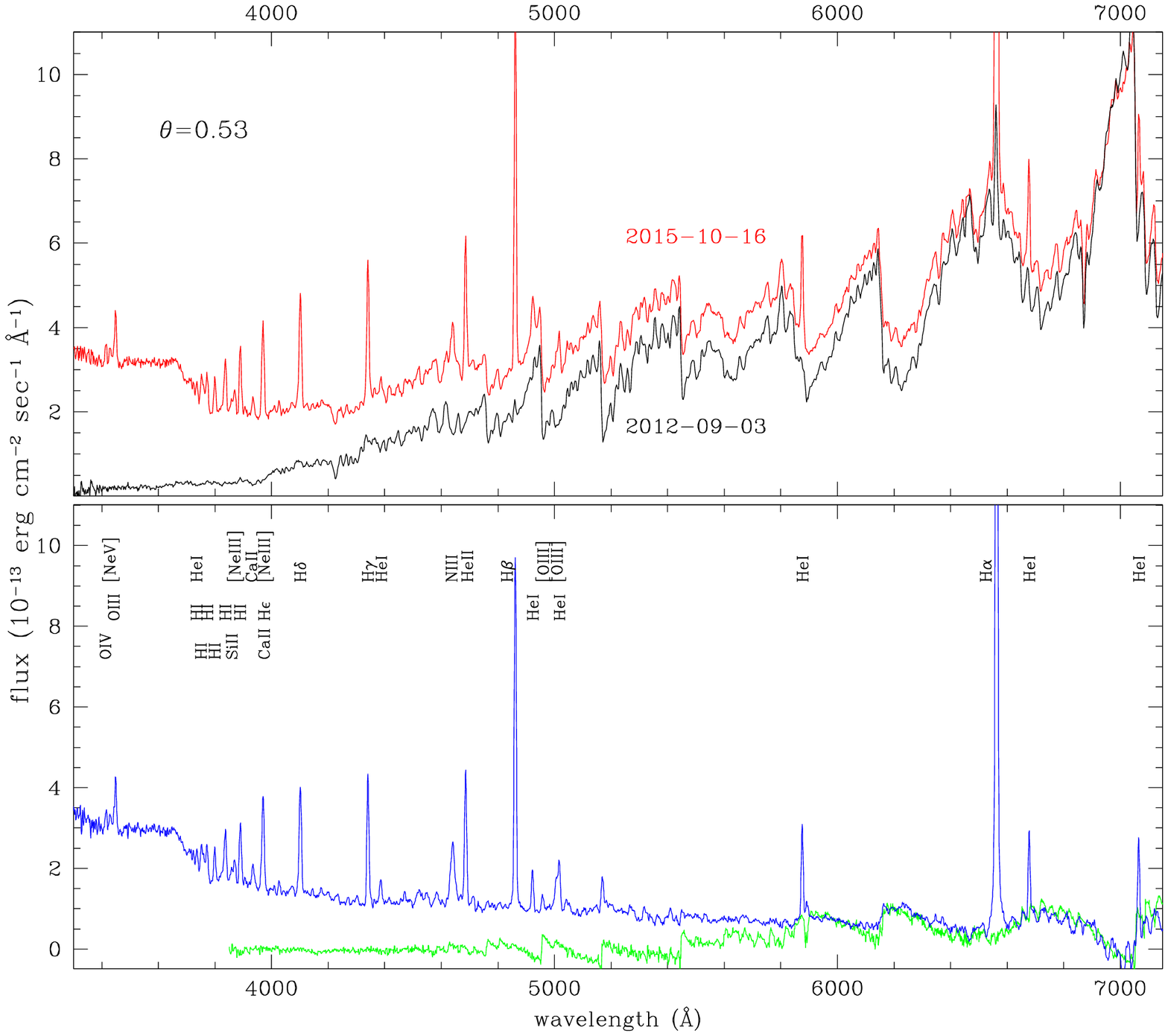}
     \caption{{\em Top:} spectra of T CrB for quiescence (2012-09-03) and
     super-active states (2015-10-16) are compared for exactly
     the same orbital phase (0.53), to cancel out any dependence from orbital
     aspect.  {\em Bottom:} their subtraction results in the nebular
     spectrum plotted here. The principal emission lines are identified and
     their integrated fluxes are listed in Table~2.}
     \label{fig5}
  \end{figure*}

  \begin{figure*}[!Ht]
     \centering
     \includegraphics[width=16cm]{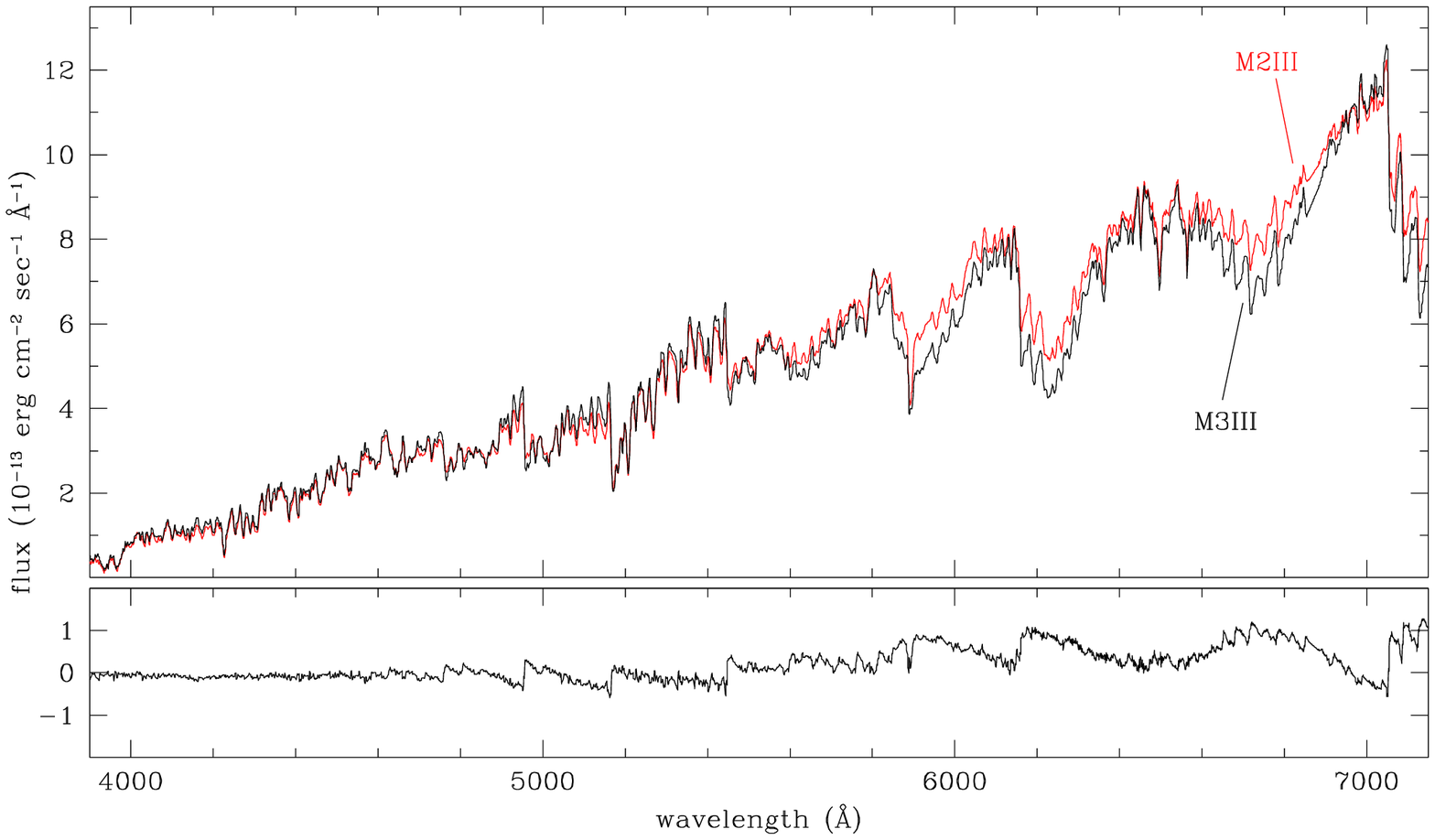}
     \caption{Spectra for M2III and M3III templates from the
     atlas of Fluks (1994), and their subtraction, showing the
     largest difference in the red.} 
     \label{fig5}
  \end{figure*}

The secular trend of T CrB in quiescence, shown in Figure~3, is
characterized by three basic features: (i) an initial faster decline, at a
mean rate of 0.0178 mag per orbital period, lasting for the initial $\sim$12
years (1947-1959), (ii) a slower decline, at a mean rate of 0.0025 mag per
orbital period, characterizing the following $\sim$55 years until present
time, and (iii) a few episodes of enhanced brightness occouring after 1975. 
Four such episodes are clearly present in Figure~3: the first three occurred
in 1975-1985, 1996-1997 and 2001-2004, and were of decreasing peak
brightness; the fourth and last one, SACT-2015 is characterized by the
largest amplitude with respect the underlying secular trend.

   \begin{table}[!Ht]
      \caption{Integrated fluxes (in units of 10$^{-13}$ erg cm$^{-2}$
      sec$^{-1}$) from Asiago 1.22m+B\&C spectra for H$\alpha$, H$\beta$,
      HeI 5876~\AA, and HeII 4686~\AA\ emission lines during the 2015
      super-active state of T CrB.}
       \centering
       \includegraphics[width=62mm]{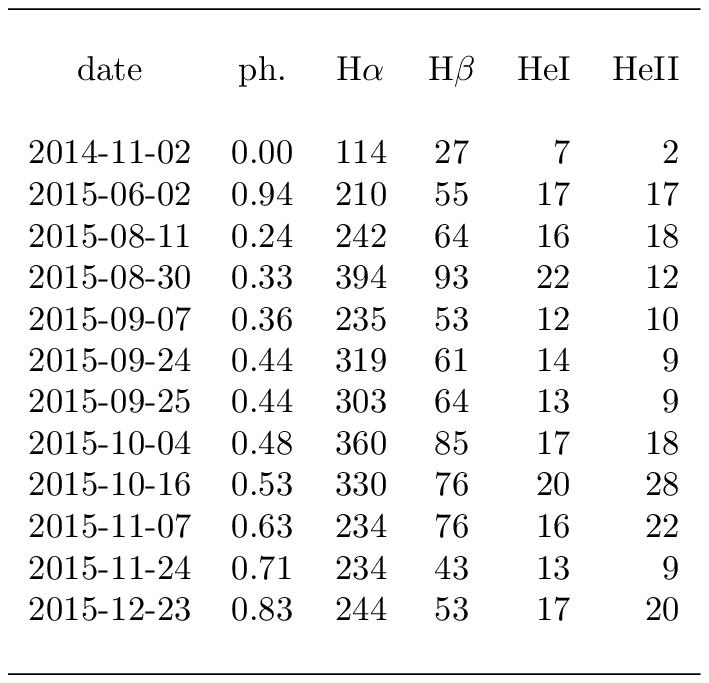}
       \label{tab3}
   \end{table}

  \begin{figure*}[!Ht]
     \centering
     \includegraphics[width=16cm]{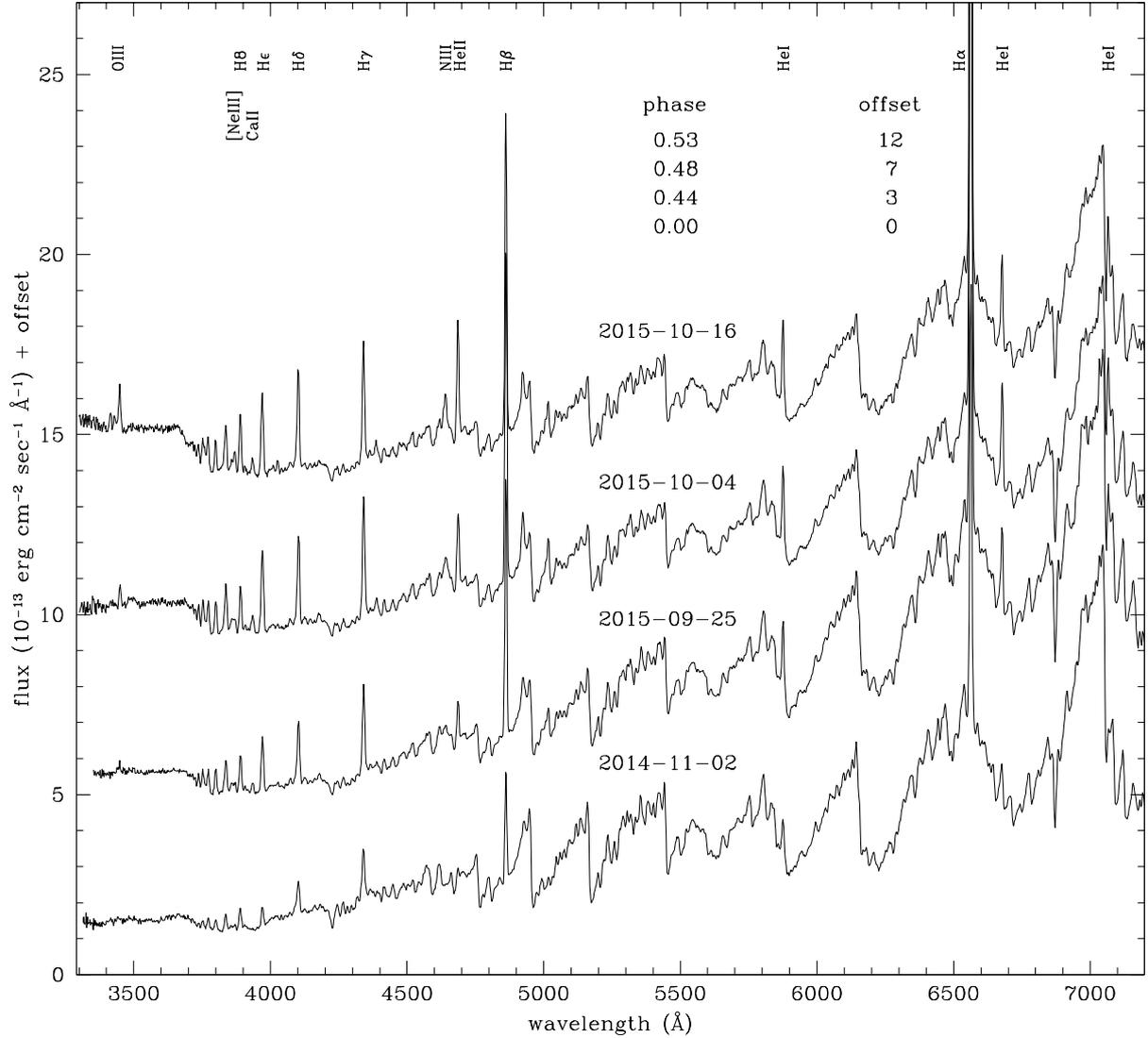}
     \caption{Sample of low res spectra of T CrB obtained during the
     2015 super-active phase, arranged in order of increasing HeII 
     4686~\AA\ flux.}
     \label{fig5}
  \end{figure*}

  \begin{figure*}[!Ht]
     \centering
     \includegraphics[angle=270,width=16.5cm]{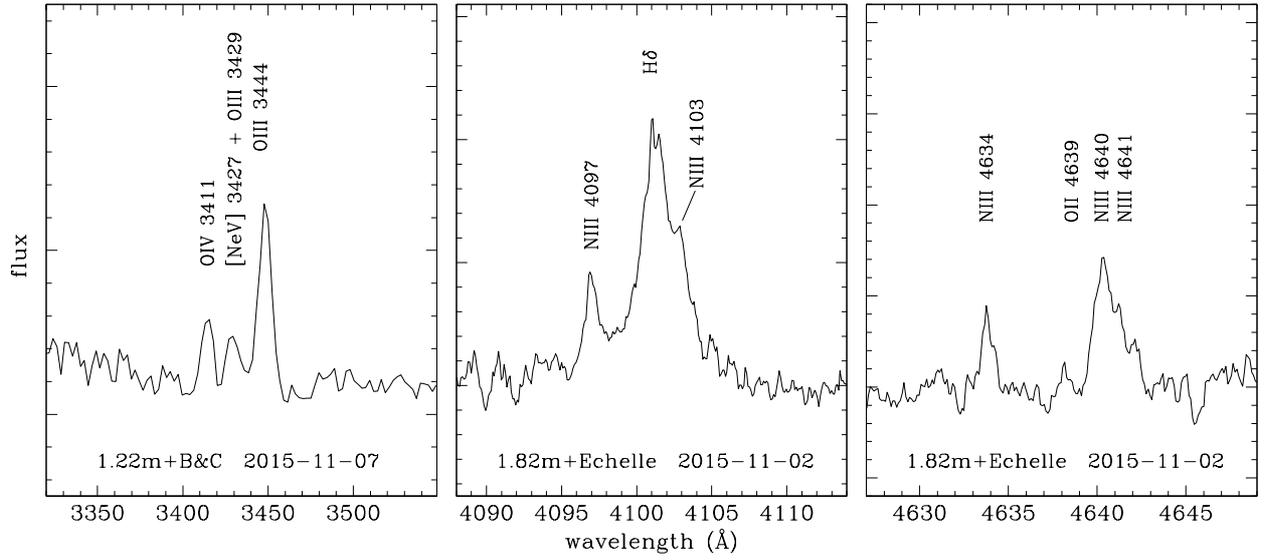}
     \caption{Zooming on different portions of the spectra of T CrB for the
     2015 super-active state, to document the presence of OIV and [NeV] in
     emission and the main OIII and NIII lines produced by the Bowen
     fluorescence mechanism.}
     \label{fig6}
  \end{figure*}

  \begin{figure*}[!Ht]
     \centering
     \includegraphics[width=16.5cm]{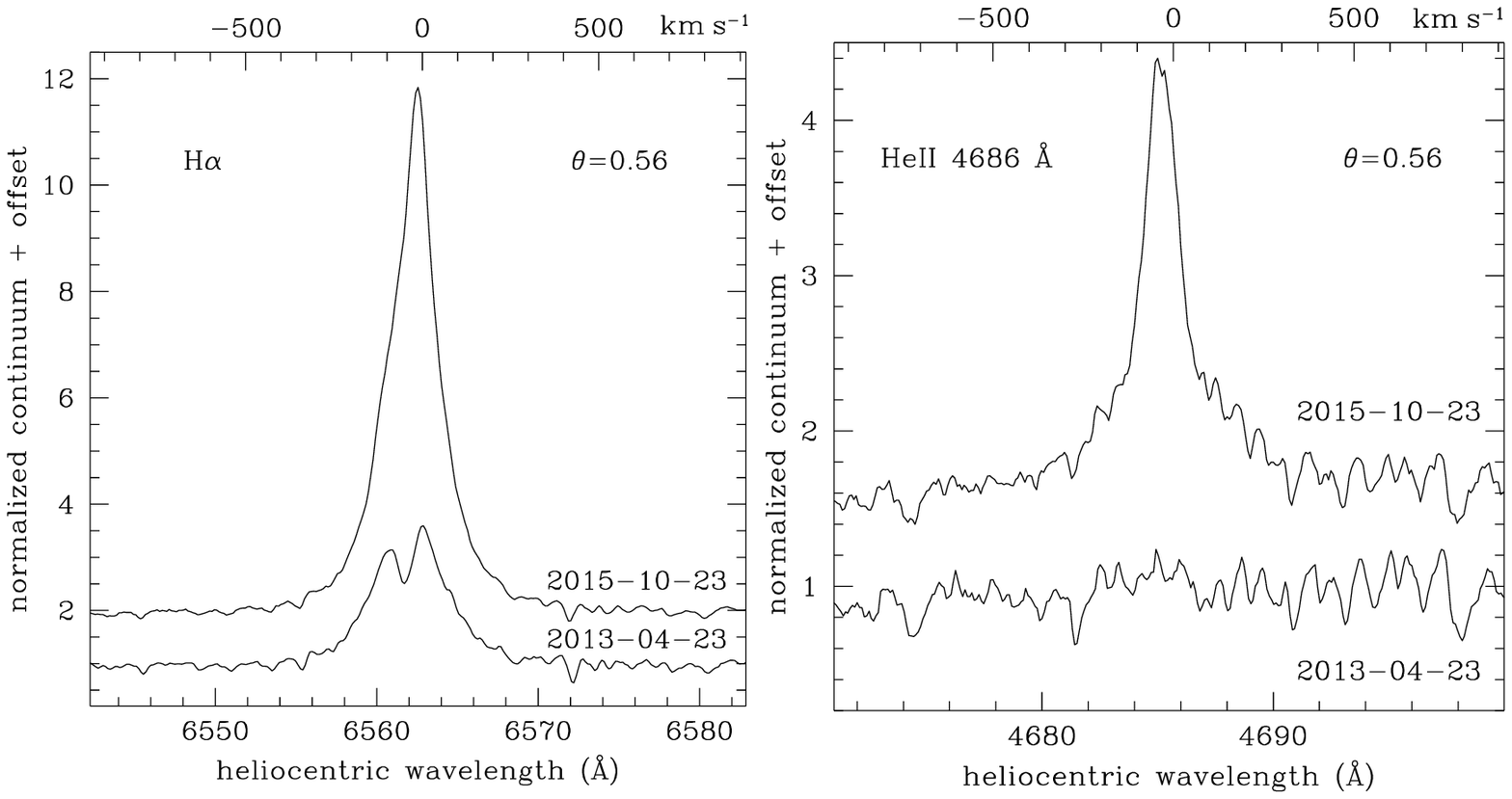}
     \caption{Comparison of the spectral appearance of T CrB around
     H$\alpha$ and HeII 4686~\AA\ during quiescence and 2015 auper-active
     state. Asiago 1.82m+Echelle spectra are compared for exactly the same
     orbital phase, to cancel out any dependence from orbital aspect.}
     \label{fig7}
  \end{figure*}

\section{Spectroscopy of the 2015 super-active state}
 
In addition to the largest increase in brightness, SACT-2015 has also seen
the greatest spectroscopic changes displayed by T CrB since the 1946 nova
outburst.  The most obvious changes are the unprecedented intensity attained
by HeII 4686 (in excess of H$\gamma$), the large intensity of OIII and NIII
lines involved in the Bowen fluorescence mechanism, and the appearance of
high ionization lines like [NeV] 3427, all these on top of strong nebular
and Balmer continua.

The spectroscopic changes are best illustrated by Figure~4, where two
spectra for exactly the same orbital phase ($\theta$=0.53, WD passing at
inferior conjunction) are compared, one from SACT-2015 and the other from
the preceeding quiescence.  The 2015-10-16 spectrum corresponds to the
strongest recorded intensity of HeII during SACT-2015, while the quiescence
2012-09-03 spectrum is well representative of the usual (and dull) apperance
of T CrB during quiescence, when only some weak emission in H$\alpha$ is
visible over an otherwise normal M3III absorption spectrum.  The result of
the subtraction of the quiescence (2012-09-03) from the SACT-2015 spectrum
(2015-10-16) is plotted on the lower panel of Figure~4.  It is a fine
example of emission from ionized gas of high density (plots built in the same
way from other SACT-2015 spectra provide similar results).  Integrated flux
of the emission lines identified in this nebular spectrum are provided in
Table~2.

At the reddest wavelengths, weak features are left in the subtraction of the
TiO molecular bands from the two spectra of Figure~4.  To evaluate them, in
Figure~5 we plot template spectra for M2III and M3III giants taken from the
atlas of Fluks et al.  (1994), after scaling them to the $V$ and $V-I_{\rm
C}$ mean values of T CrB in quiescence at the same orbital phase
($\theta$=0.53).  In the same Figure~5 we plot their difference, which is also
over-plotted to the nebular spectrum of Figure~4, where it provides a
perfect match to the weak features left over at reddest wavelengths by the
subtraction of the SACT-2015 and quiescence spectra.  This indicates that
the irradiation by the hot source has rised the temperature of the facing side of
the giant companion, from that of a M3III during quiescence to that of a
M2III during SACT-2015.  This change corresponds to an increase in the
effective temperature of $\Delta$$T_{\rm eff}$$\sim$80~K, averaging between
$\Delta$$T_{\rm eff}$=90~K and $\Delta$$T_{\rm eff}$=70~K reported by
Ridgway et al.  (1980) and Fluks et al.  (1994), respectively, as the
difference in temperature between M2III and M3III giants.

The intensity of HeII has varied considerably during SACT-2015, as
illustrated by the sample of spectra presented in Figure~6.  In this figure
we also plot for comparison the spectrum for 2014-11-02, that caught T CrB on
the transition from quiescence to SACT-2015, which show increased nebular
emission (both continuum and lines) and the first appearance of HeII.
Table~3 lists the integrated flux for a few representative lines as measured
on our SACT-2015 low resolution spectra.

\section{The nebular spectrum}

Some comments are here in order concerning the nebular spectrum of T CrB
during SACT-2015, a detailed photo-ionization modeling being pursued
elsewhere.

The fluxes of emission lines listed in Tables~2 and 3 are little affected by
the low extinction experienced by T CrB.  The 3D Galactic dust model by
Munari et al.  (2014) indicate a total interstellar extinction
$E_{B-V}$=0.048 along the line of sight to T CrB, and similarly very low
values of $E_{B-V}$=0.058 and $E_{B-V}$=0.067 are derived from the 3D
Galactic dust distributions of Schlegel, Finkbeiner, \& Davis (1998) and Schlafly
\& Finkbeiner (2011), respectively.  Cassatella et al.  (1982) from analysis
of IUE ultraviolet spectra derived a slightly larger reddening,
$E_{B-V}$=0.15, that could indicate some contribution by local
circumstellar matter around T CrB.  After dereddening, the ratio of Balmer
line fluxes listed in Table~2 suggests a negligible optical depth in
H$\alpha$, following the analysis of Barker (1978) and Feibelman (1983).

The spectra of T CrB in Figure~4 display a strong $\lambda$ 4640~\AA\ blend. 
It is due to three NIII lines (multiplet N.2) pumped by Bowen fluorescence
mechanism (BFM in the following; Bowen 1934, 1935).  These lines (4634,
4640, 4641~\AA) are resolved in the Echelle spectrum for 2015-11-02
presented in Figure~7 (right panel).  Their production begins with emission
of HeII Ly-$\alpha$ photons at 303.8~\AA, which wavelength corresponds to
that of OIII transiting from its ground state to an excited level.  The OIII
downward transitions produce photons at 3415, 3428 and 3444~\AA\, in the
1:8:54 proportions.  The OIII 3444 line is observed in strong emission in
SACT-2015 spectra, as illustrated in the left panel of Figure~7, while the
other two OIII lines are blended with nearby lines.  The end product of the
OIII downward transitions is the emission of photons at 374.4~\AA\ which
correspond to the transition of NIII from its ground to an excited state. 
The following de-excitation results in the emission of the three lines
constituting the $\lambda$ 4640~\AA\ blend above mentioned, and in a pair of
lines at 4097 and 4103~\AA\ (NIII multiplet N.1), which are also in strong
emission in T CrB as illustrated by the central panel of Figure~7.  It is
also worth noticing that in the SACT-2015 spectra Figure~6, the intensity of
OIII 3444 and NIII 4640 blend varies in parallel with that of HeII 4686, as
expected when BFM is ruling.  The BFM pumping has been studied in detail in
symbiotic stars by Eriksson et al.  (2005) and Selvelli et al.  (2007).  The
efficency of the BFM in T CrB during SACT-2015 (defined as the fraction of
HeII Ly-$\alpha$ photons that leads to OIII upward transitions) is
$\sim$0.35, following Harrington (1972) formalism.

We have detected [NeV] 3427 emission line in the optical spectra of T CrB,
to the best of our knowledge the first time this have occourred away from
nova eruptions, a further indication of the exceptional state T CrB has
underwent in 2015.  The line is identified in the portion of T CrB
spectrum highlighted in the left panel of Figure~7.  The line is observed at
3428~\AA, at the mean position for [NeV] 3427 and OIII 3429, which
contributes equal amounts to the observed flux.  In fact, from Table~2, the
observed flux for the 3428 blend is 1/3 of OIII 3444, while that expected
from OIII 3429 alone would be 1/6.5, according to both theoretical
transition probabilities and actual observations in symbiotic binaries
(Selvelli et al.  2007). 

The simultaneous presence of both [NeV] 3427 and [NeIII] 3869 could suggest
that [NeIV] lines should be equally present in the nebular spectrum of T
CrB.  The strongest optical [NeIV] lines are located at 4715-4727~\AA\
(Merrill 1956), and they never attend a significant intensity (when seen in
symbiotic binaries, they score only a few \% of the intensity of [NeV] and
[NeIII] lines; cf.  Allen 1983, Munari \& Zwitter 2002).  The non-detection
of [NeIV] in the spectra of T CrB is therefore not surprising.  Similarly,
the strongest OIV line that is observed at optical wavelengths is the 3411
\AA\ (Jaschek \& Jaschek 2009), and given its modest intensity in the
spectra of T CrB no further lines from this ion are expected to be visible.

The relative intensity of HeI and HeII emission lines varied greatly during
SACT-2015, as illustrated by the sample of spectra plotted in Figure~6 and
the line fluxes listed in Table~3.  The HeII 4686 / (HeI 5876 + HeI 6678)
ratio is seen to vary from 0.17 on 2014-11-02, to 0.34 on 2015-08-30, to
0.88 on 2015-10-16.  For the latter two dates, the HeI lines augment their
intensity by just 20\% while HeII increases by three times.  This behaviour
indicates that during quiescence and the initial rise toward SACT-2015, the
nebula was {\em ionization bounded}, with properly nested Stromgren's
spheres for different ions and neutral material further out, ready to be
ionized by an increase in the hot source output.  During SACT, the nebula became
{\em density bounded}, i.e.  all the available gas was already ionized and
any increase in the hot source output could only rise the ionization degree of
the gas but not further expand the nebula into pre-existing external
neutral material.  The ionization to density bounded transition is nicely
confirmed by the evolution of the H$\alpha$ profiles shown in Figure~8,
where a quiescence (2013-04-23) and a SACT-2015 (2015-10-26) profile for
exactly the same orbital phase (0.56) are compared.  The H$\alpha$ profile
for quiescence shows a weak emission and superimposed to it a narrow
aborption, which is missing from the SACT-2015 profile that displays a
vastly stronger emission.  This narrow absorption is typical of symbiotic
stars, and originates in the outflowing wind of the cool giant, specifically
from the neutral portion external to the fraction ionized by the WD, as it
was demonstrated by Munari (1993) who followed for several cycles the
orbital motion of the cool giant, of the emission lines and of the narrow
absorptions in EG And, a symbiotic star with optical spectra closely similar
to those of T CrB in quiescence.  The absence of the narrow central
aborption from the SACT-2015 profile indicate that, in the direction of the
observer, no neutral gas exists external to the ionized gas.  The velocity
of the narrow absorption is -19 km~sec$^{-1}$ with respect to that of the
cool giant, which is therefore the terminal velocity of its outfowing wind,
a value typical of cool giants.

Finally, the values reported in Table~3 shows how the intensity of HeII 4686
emission line is more responsive to the varying activity of the hot source than to
the orbital aspect.

\section{Three levels of activity for T CrB in quiescence}

Iijima (1990) noted that during quiscence, i.e. away from the 1866 and 1946
nova outbursts, T CrB exhibits two states: an ``high" one when emission
lines (Balmer, HeI) and the nebular continuum are relatively strong, and a
``low" state when they essentially disappear (except some weak residual
emission in H$\alpha$). The unique conditions eperienced by T CrB during
SACT-2015 requires the introduction of a new, third state that we term
``super-active", which is characterized by (1) the presence of OIV and [NeV]
lines and a very strong HeII 4686, a strong 4640 Bowen fluorescence blend,
(2) a large increse in mean brightness, and (3) disappearance of orbital
modulation from $B$-band lightcurve.

We have searched the available leterature (e.g. Kraft 1958, Gravina 1981,
Andrillat \& Houziaux 1982, Blair et al.  1983, Williams 1983, Kenyon and
Garcia 1986, Iijima 1990, Anupama \& Prabhu 1991, Ivison et al.  1994,
Anupama 1997, Zamanov and Marti 2001, Munari \& Zwitter 2002) in the attempt
to reconstruct the history of spectroscopic activity of T CrB during
quiescence.  This has turned out a difficult task because rarely integrated
absolute fluxes are provided for the emission lines, few observations
ventured enough into the blue to cover the Balmer continuum, and usually
only equivalent widths are given if not just a mere descrition like 'weak'
or 'strong'.  In addition the observations reported in literature were
obtained at different wavelength intervals and resolving powers.  We tried
our best to homogenize the different sources, and in this we took advantage
of the many (unpublished) spectra of T CrB that we have regularly obtained
since 1987.

T CrB has always been in a 'low' state when observed for the first 3 decades
after the 1946 outburst.  The last of these spectra, those of Blair et al. 
(1983) for 1981-02-06, Williams (1983) for 1981-06-10, and Gravina (1981)
for 1981-07-15 and 1981-09-15 record only feeble emission in H$\alpha$,
H$\beta$ and H$\gamma$.  Iijima (1990) reports that, in addition to Balmer
and HeI, a weak emission in HeII 4686~\AA\ was visible on his spectra on
several dates distributed between 1982 and 1987, but Kenyon and Garcia
(1986) saw no HeII in emission on their 1984 and 1985 spectra and the
absolute flux they measured for Balmer lines was only twice larger than that
reported by Blair et al.  (1983) for 1981.  The extensive spectral
monitoring by Anupama \& Prabhu (1991) and Anupama (1997) shows that the
intensity of Balmer and HeI emission lines gradually increased starting with
June 1985, peaked during November 1986, and returned to the 'low' state by
October 1987, where T CrB has remained until 1996.  Iijima (1990) confirms
that HeII was absent from his spectra for 1988, 1989 and 1990, the same
reported by Ivison et al (1994) for their 1989 spectra. Just a feable
emission in the lower Balmer lines and no HeII were found by Munari \&
Zwitter (2002) on various dates of 1993 and 1995. Then a new 'high' state
was briefly observed in 1996-1997. On 1996-02-01 Zamanov \& Marti (2001)
found H$\alpha$ to be weak and this is confirmed by a 1996-02-08 spectrum
from Munari \& Zwitter (2002) that in addition reveals HeII to be absent.
Then, Mikolajewski et al. (1997) found H$\alpha$ to be in strong emission
during April, May and June of 1996. This is confirmed by a 1996-05-30
spectrum from Munari \& Zwitter (2002), that in addition shows how HeII was
still absent. Zamanov \& Marti (2001) reports that by 1998 this second
'high' state of T CrB was over. Since then and to the best of our knowledge,
T CrB has never been observed again in a 'high' state until the 2015 episode
described in this paper.  The long term photometric behaviour presented in
Figure~3 shows that T CrB rised significantly above the underlying secular
decline only in correspondence of the high spectroscopic states.

To the best of our knowledge, T CrB has been caught in a super-active state
in only one other occasion, on the summer of 1938 by Hachenberg \& Wellmann
(1939). On their spectrum for 22 July 1938, HeII 4686 is half the intensity
of H$\gamma$ and the 4640~\AA\ blend stands in prominent emission (2/3 the
intensity of HeII). The Hachenberg \& Wellmann (1939) spectrum for August 28
confirms the super-active state, while that for September 22 indicates a
rapid return of T CrB toward lower excitation conditions.

There is an intriguing parallelism between SACT-2015 and what Hachenberg \&
Wellmann (1939) observed in 1938. The super-active state they caught
occurred $\sim$70 years past the 1866 nova outburst, and SACT-2015 is
occurring $\sim$70 years past the 1946 nova outburst. Is therefore everything
in place for a new nova outburst in 2026, again $\sim$80 years past the last
eruption~?



\begin{thebibliography}{00}
 \bibitem[\protect\citeauthoryear{Allen}{1983}]{1983MNRAS.204..113A} Allen D.~A., 1983, MNRAS, 204, 113 
 \bibitem[\protect\citeauthoryear{Allen}{1984}]{1984PASAu...5..369A} Allen D.~A., 1984, PASAu, 5, 369 
 \bibitem[\protect\citeauthoryear{Andrillat  \& Houziaux}{1982}]{1982ASSL...95...57A} Andrillat Y., Houziaux L., 1982, in The nature of symbiotic stars, M. Fridjung and R. Viotti eds., Reidel ASSL 95, 57 
 \bibitem[\protect\citeauthoryear{Anupama}{1997}]{1997ppsb.conf..117A} Anupama G.~C., 1997, in Physical Processes in Symbiotic Binaries and Related Systems, J. Mikolajewska ed.,                                  Copernicus Foundation for Polish Astronomy, 117 
 \bibitem[\protect\citeauthoryear{Anupama \& Prabhu}{1991}]{1991MNRAS.253..605A} Anupama G.~C., Prabhu T.~P., 1991, MNRAS, 253, 605 
 \bibitem[\protect\citeauthoryear{Bailey}{1975}]{1975JBAA...85..217B} Bailey J., 1975, JBAA, 85, 217 
 \bibitem[\protect\citeauthoryear{Barnard}{1907}]{1907ApJ....25..279B} Barnard E.~E., 1907, ApJ, 25, 279 
 \bibitem[\protect\citeauthoryear{Belczynski \& Mikolajewska}{1998}]{1998MNRAS.296...77B} Belczynski K., Mikolajewska J., 1998, MNRAS, 296, 77 
 \bibitem[\protect\citeauthoryear{Blair et al.}{1983}]{1983ApJS...53..573B}  Blair W.~P., Feibelman W.~A., Michalitsianos A.~G., Stencel R.~E., 1983,  ApJS, 53, 573 
 \bibitem[\protect\citeauthoryear{Bowen}{1934}]{1934PASP...46..146B} Bowen I.~S., 1934, PASP, 46, 146 
 \bibitem[\protect\citeauthoryear{Bowen}{1935}]{1935ApJ....81....1B} Bowen I.~S., 1935, ApJ, 81, 1 
 \bibitem[\protect\citeauthoryear{Campbell \& Shapley}{1923}]{1923HarCi.247....1C} Campbell L., Shapley H., 1923, HarCi, 247, 1 
 \bibitem[\protect\citeauthoryear{Cannizzo \& Kenyon}{1992}]{1992ApJ...386L..17C} Cannizzo J.~K., Kenyon S.~J., 1992, ApJ, 386, L17 
 \bibitem[\protect\citeauthoryear{Cassatella et al.}{1982}]{1982ESASP.176..229C} Cassatella A., Patriarchi P., Selvelli P.~L., Bianchi L., Cacciari C., Heck A., Perryman M., Wamsteker W., 1982, ESASP, 176, 229 
 \bibitem[\protect\citeauthoryear{Dobrotka et al.}{2010}]{2010MNRAS.402.2567D} Dobrotka A., Hric L., Casares J., Shahbaz T., Mart{\'{\i}}nez-Pais I.~G., Mu{\~n}oz-Darias T., 2010, MNRAS, 402, 2567 
 \bibitem[\protect\citeauthoryear{Eriksson et al.}{2005}]{2005A&A...434..397E} Eriksson M., Johansson S., Wahlgren G.~M., Veenhuizen H., Munari U., Siviero A., 2005, A\&A, 434, 397 
 \bibitem[\protect\citeauthoryear{Feibelman}{1983}]{1983ApJ...275..628F} Feibelman W.~A., 1983, ApJ, 275, 628 
 \bibitem[\protect\citeauthoryear{Fekel et al.}{2000}]{2000AJ....119.1375F} Fekel F.~C., Joyce R.~R., Hinkle K.~H., Skrutskie M.~F., 2000, AJ, 119, 1375 
 \bibitem[\protect\citeauthoryear{Fluks et al.}{1994}]{1994A&AS..105..311F} Fluks M.~A., Plez B., The P.~S., de Winter D., Westerlund B.~E., Steenman H.~C., 1994, A\&AS, 105, 311 
 \bibitem[\protect\citeauthoryear{Gravina}{1981}]{1981IBVS.2041....1G} Gravina R., 1981, IBVS, 2041, 1 
 \bibitem[\protect\citeauthoryear{Gromadzki et al.}{2006}]{2006AcA....56...97G} Gromadzki M., Mikolajewski M., Tomov T., Bellas-Velidis I., Dapergolas A., Galan C., 2006, AcA, 56, 97 
 \bibitem[\protect\citeauthoryear{Hachenberg \& Wellmann}{1939}]{1939ZA.....17..246H} Hachenberg O., Wellmann P., 1939, ZA, 17, 246 
 \bibitem[\protect\citeauthoryear{Harrington}{1972}]{1972ApJ...176..127H} Harrington J.~P., 1972, ApJ, 176, 127 
 \bibitem[\protect\citeauthoryear{Henden \& Munari}{2006}]{2006A&A...458..339H} Henden A., Munari U., 2006, A\&A, 458, 339 
 \bibitem[\protect\citeauthoryear{Iijima}{1990}]{1990JAVSO..19...28I} Iijima T., 1990, JAVSO, 19, 28 
 \bibitem[\protect\citeauthoryear{Ivison, Bode, \& Meaburn}{1994}]{1994A&AS..103..201I} Ivison R.~J., Bode M.~F., Meaburn J., 1994, A\&AS, 103, 201 
 \bibitem[\protect\citeauthoryear{Kastner \& Bhatia}{1996}]{1996MNRAS.279.1137K} Kastner S.~O., Bhatia A.~K., 1996, MNRAS, 279, 1137 
 \bibitem[\protect\citeauthoryear{Kenyon}{1986}]{1986syst.book.....K} Kenyon S.~J., 1986, The Symbiotic Stars, Cambridge Univ. Press  
 \bibitem[\protect\citeauthoryear{Kenyon \& Garcia}{1986}]{1986AJ.....91..125K} Kenyon S.~J., Garcia M.~R., 1986, AJ, 91, 125 
 \bibitem[\protect\citeauthoryear{Jaschek \& Jaschek}{2009}]{2009bces.book.....J} Jaschek C., Jaschek M., 2009, The Behavior of Chemical Elements in Stars, Cambridge University Press  
 \bibitem[\protect\citeauthoryear{Merrill}{1956}]{1956lcea.book.....M} Merrill P.~W., 1956, Lines of the Chemical Elements in Astronomical Spectra, Carnegie Institution of Washington Publication 610  
 \bibitem[\protect\citeauthoryear{Mikolajewski, Tomov, \& Kolev}{1997}]{1997IBVS.4428....1M} Mikolajewski M., Tomov T., Kolev D., 1997, IBVS, 4428, 1 
 \bibitem[\protect\citeauthoryear{Munari}{1993}]{1993A&A...273..425M} Munari U., 1993, A\&A, 273, 425
 \bibitem[\protect\citeauthoryear{Munari}{1997}]{1997ppsb.conf...37M} Munari U., 1997, in Physical Processes in Symbiotic Binaries and Related Systems, J. Mikolajewska ed.,                                  Copernicus Foundation for Polish Astronomy, p.37 
 \bibitem[\protect\citeauthoryear{Munari et al.}{2012}]{2012BaltA..21...13M} Munari U., et al., 2012, BaltA, 21, 13 
 \bibitem[\protect\citeauthoryear{Munari et al.}{2014}]{2014AJ....148...81M} Munari U., et al., 2014, AJ, 148, 81 
 \bibitem[\protect\citeauthoryear{Munari \& Moretti}{2012}]{2012BaltA..21...22M} Munari U., Moretti S., 2012, BaltA, 21, 22 
 \bibitem[\protect\citeauthoryear{Munari \& Zwitter}{2002}]{2002A&A...383..188M} Munari U., Zwitter T., 2002, A\&A, 383, 188 
 \bibitem[\protect\citeauthoryear{Osterbrock \& Ferland}{2006}]{2006agna.book.....O} Osterbrock D.~E., Ferland G.~J., 2006, Astrophysics of gaseous nebulae and active galactic nuclei, 2nd. ed., University Science Books
 \bibitem[\protect\citeauthoryear{Paine-Gaposchkin}{1957}]{1957gano.book.....G} Paine-Gaposchkin C.~H., 1957, The Galactic Novae, North-Holland Pub. Co.  
 \bibitem[\protect\citeauthoryear{Pettit}{1946}]{1946PASP...58..153P} Pettit E., 1946, PASP, 58, 153 
 \bibitem[\protect\citeauthoryear{Ridgway et al.}{1980}]{1980ApJ...235..126R} Ridgway S.~T., Joyce R.~R., White N.~M., Wing R.~F., 1980, ApJ, 235, 126 
 \bibitem[\protect\citeauthoryear{Ruffert, Cannizzo, \& Kenyon}{1993}]{1993ApJ...419..780R} Ruffert M., Cannizzo J.~K., Kenyon S.~J., 1993, ApJ, 419, 780 
 \bibitem[\protect\citeauthoryear{Sanford}{1946}]{1946PASP...58..240S} Sanford R.~F., 1946, PASP, 58, 240 
 \bibitem[\protect\citeauthoryear{Sanford}{1947}]{1947PASP...59...87S} Sanford R.~F., 1947, PASP, 59, 87 
 \bibitem[\protect\citeauthoryear{Sanford}{1949}]{1949ApJ...109...81S} Sanford R.~F., 1949, ApJ, 109, 81 
 \bibitem[\protect\citeauthoryear{Schaefer}{2010}]{2010ApJS..187..275S} Schaefer B.~E., 2010, ApJS, 187, 275 
 \bibitem[\protect\citeauthoryear{Schlafly \& Finkbeiner}{2011}]{2011ApJ...737..103S} Schlafly E.~F., Finkbeiner D.~P., 2011, ApJ, 737, 103 
 \bibitem[\protect\citeauthoryear{Schlegel, Finkbeiner, \& Davis}{1998}]{1998ApJ...500..525S} Schlegel D.~J., Finkbeiner D.~P., Davis M., 1998, ApJ, 500, 525 
 \bibitem[\protect\citeauthoryear{Selvelli, Cassatella, \& Gilmozzi}{1992}]{1992ApJ...393..289S} Selvelli P.~L., Cassatella A., Gilmozzi R., 1992, ApJ, 393, 289 
 \bibitem[\protect\citeauthoryear{Selvelli, Danziger, \& Bonifacio}{2007}]{2007A&A...464..715S} Selvelli P., Danziger J., Bonifacio P., 2007, A\&A, 464, 715 
 \bibitem[\protect\citeauthoryear{Skopal}{2005}]{2005A&A...440..995S} Skopal A., 2005, A\&A, 440, 995 
 \bibitem[\protect\citeauthoryear{Warner}{1995}]{1995CAS....28.....W} Warner B., 1995, Cataclysmic Variable Stars, Cambridge Astrophysics Series, 28,  
 \bibitem[\protect\citeauthoryear{Webbink}{1976}]{1976Natur.262..271W} Webbink R.~F., 1976, Natur, 262, 271 
 \bibitem[\protect\citeauthoryear{Williams}{1983}]{1983ApJS...53..523W}  Williams G., 1983, ApJS, 53, 523 
 \bibitem[\protect\citeauthoryear{Williams}{1992}]{1992AJ....104..725W} Williams R.~E., 1992, AJ, 104, 725 
 \bibitem[\protect\citeauthoryear{Yudin \& Munari}{1993}]{1993A&A...270..165Y} Yudin B., Munari U., 1993, A\&A, 270, 165 
 \bibitem[\protect\citeauthoryear{Zamanov \& Bruch}{1998}]{1998A&A...338..988Z} Zamanov R.~K., Bruch A., 1998, A\&A, 338, 988 
 \bibitem[\protect\citeauthoryear{Zamanov et al.}{2004}]{2004MNRAS.350.1477Z} Zamanov R., Bode M.~F., Stanishev V., Mart{\'{\i}} J., 2004, MNRAS, 350, 1477 
 \bibitem[\protect\citeauthoryear{Zamanov \& Marti}{2001}]{2001IBVS.5013....1Z} Zamanov R., Marti J., 2001, IBVS, 5013, 1 
\end{thebibliography}
\end{document}